\documentclass[prl,final,twocolumn,showpacs,floatfix]{revtex4}%
\usepackage{amsfonts}
\usepackage{amsmath}
\usepackage{amssymb}
\usepackage{graphicx}%
\providecommand{\U}[1]{\protect\rule{.1in}{.1in}}

\begin{document}
\title{\textit{Ab initio} calculation of the Hoyle state}
\author{Evgeny Epelbaum$^{a}$, Hermann~Krebs$^{a}$, Dean~Lee$^{b}$, Ulf-G.~Mei{\ss }%
ner$^{c,d}$}
\affiliation{$^{a}$Institut f\"{u}r Theoretische Physik II, Ruhr-Universit\"{a}t Bochum,
D-44870 Bochum, Germany \linebreak$^{b}$Department of Physics, North Carolina
State University, Raleigh, NC 27695, USA \linebreak$^{c}$Helmholtz-Institut
f\"{u}r Strahlen- und Kernphysik and Bethe Center for Theoretical Physics,
Universit\"{a}t Bonn, D-53115 Bonn, Germany \linebreak$^{d}$Institut f\"{u}r
Kernphysik, Institute for Advanced Simulation and J\"{u}lich Center for Hadron
Physics, Forschungszentrum J\"{u}lich, D-52425 J\"{u}lich, Germany }

\begin{abstract}
The Hoyle state plays a crucial role in the hydrogen burning of stars heavier
than our sun and in the production of carbon and other elements necessary for
life. \ This excited state of the carbon-12 nucleus was postulated by Hoyle
\cite{1} as a necessary ingredient for the fusion of three alpha particles to
produce carbon at stellar temperatures. \ Although the Hoyle state was seen
experimentally more than a half century ago \cite{2,3} nuclear theorists have
not yet uncovered the nature of this state from first principles. \ In this
letter we report the first \textit{ab initio} calculation of the low-lying
states of carbon-12 using supercomputer lattice simulations and a theoretical
framework known as effective field theory. \ In addition to the ground state
and excited spin-2 state, we find a resonance at $-85(3)$~MeV with all of the
properties of the Hoyle state and in agreement with the experimentally
observed energy. \ These lattice simulations provide insight into the
structure of this unique state and new clues as to the amount of fine-tuning
needed in nature for the production of carbon in stars.

\end{abstract}

\maketitle

In stars with central temperatures above $15\times10^{6}$~K, the
carbon-nitrogen-oxygen cycle is the dominant process for the conversion of
hydrogen into helium \cite{4,5}. \ However a key catalyst in this cycle is the
carbon-12 nucleus which itself must be produced by fusion of three helium-4
nuclei or alpha particles. \ Without additional help this triple alpha
reaction is highly suppressed at stellar temperatures and presents a
bottleneck shutting down other process. \ Fortunately several coincidences
prevent this from happening. \ The first stage fusing together two alpha
particles is enhanced by the beryllium-8 ground state, a resonance very near
the double alpha threshold. \ In order to enhance the fusion of the third
alpha particle, Hoyle postulated a new excited state of $^{12}$C, a spinless
even-parity resonance very near the $^{8}$Be-alpha threshold \cite{1}. \ Soon
after this prediction, the state was found at Caltech \cite{2,3} and has been
investigated in laboratories worldwide. \ Given its role in the formation of
life-essential elements, this state is commonly mentioned in anthropic
arguments explaining the fine-tuning of fundamental parameters of the universe
\cite{6}.

The Hoyle state presents a major challenge for nuclear theory. \ There have
been recent studies of carbon-12 and the Hoyle state built from clusters of
alpha particles \cite{7,8,9}. \ While these empirical models provide
qualitative insights, investigations of the fundamental properties of the
Hoyle state require calculations from first principles. \ One very interesting
calculation is based on fermionic molecular dynamics, but it requires a fit to
properties of a broad range of nuclei to pin down the various model parameters
\cite{9}. \ In recent years several \textit{ab initio} approaches have been
used to calculate the binding, structure, and reactions of atomic nuclei.
\ These include the no-core shell model \cite{10,11}, constrained-path Green's
function Monte Carlo \cite{12,13}, auxiliary-field diffusion Monte Carlo
\cite{14}, and coupled cluster methods \cite{15}. \ Despite spectacular
progress over the past few years, there have been no calculations so far which
reproduce the Hoyle state from first principles.

In this letter we report new \textit{ab initio }calculations of the low-lying
spectrum of carbon-12 using the framework of chiral effective field theory and
Monte Carlo lattice simulations. \ Effective field theory (EFT) is an
organizational tool which reconstructs the interactions of particles as a
systematic expansion in powers of particle momenta. \ Initiated by Weinberg in
1991 \cite{16}, chiral EFT provides a systematic hierarchy of the forces among
protons and neutrons. \ This approach comes with an estimate of the
theoretical uncertainty at any given order which can be systematically reduced
at higher orders. \ Over the past two decades, chiral EFT has proven a
reliable and precise tool to describe the physics of few-nucleon systems. \ A
recent review can be found in Ref.~\cite{17}.\ \ The low-energy expansion of
EFT is organized in powers of $Q$, where $Q$ denotes the typical momentum of
particles. \ The momentum scale $Q$ is also roughly the same size as the mass
of the pion times the speed of light. \ The most important contributions come
at leading order (LO) or $O(Q^{0})$. \ The next most important terms are at
next-to-leading order (NLO) or $O(Q^{2})$. \ The terms just beyond this are
next-to-next-to-leading order (NNLO) or $O(Q^{3})$. \ In the lattice
calculations presented here, we consider all possible interactions up to
$O(Q^{3})$. \ We also separate out explicitly the $O(Q^{2})$ terms which arise
from electromagnetic interactions (EM) and isospin symmetry breaking (IB) due
to mass differences of the up and down quarks.

Lattice effective field theory combines EFT with numerical lattice methods in
order to investigate larger systems. \ Space is discretized as a periodic
cubic lattice with spacing $a$ and length $L$, where $L$ is typically
$\sim10\,$fm. \ In the time direction, the time step is denoted $a_{t}$ with
total propagation time $L_{t}$. \ On this spacetime lattice, nucleons are
point-like particles on lattice sites. \ Interactions due to the exchange of
pions and multi-nucleon operators are generated using auxiliary fields.
\ Lattice EFT was originally used to calculate the properties of homogeneous
nuclear and neutron matter \cite{18,19}. \ Since then the ground state
energies of atomic nuclei with up to twelve nucleons have been investigated
\cite{20,21}. \ A recent review of the literature can be found in
Ref.~\cite{22}.

In the lattice calculations presented here we use the low-energy filtering
properties of Euclidean time propagation. \ If $H$ is the Hamiltonian operator
for a quantum system, then the eigenvalues of $H$ are the energy levels and
the eigenvectors of $H$ are the corresponding wavefunctions. For any given
quantum state, $\Psi$, the projection amplitude $Z_{\Psi}(t)$ is defined as
the expectation value $\left\langle e^{-Ht}\right\rangle _{\Psi}$. \ For large
Euclidean time $t$, the exponential operator $e^{-Ht}$ enhances the signal of
low-energy states. \ The corresponding energies can be determined from the
exponential decay of these projection amplitudes.

In Table~\ref{ground_states} we present lattice results for the ground state
energies of $^{4}$He$,$ $^{8}$Be, and $^{12}$C. \ The method of calculation is
essentially the same as that described in Ref.~\cite{21}. \ We note that
higher-order corrections are computed using perturbation theory. \ Some
improvements have been made which eliminate the problem of overbinding found
in Ref.~\cite{21}. \ One significant improvement involves choosing local
two-derivative lattice operators at NLO which prevent interactions tuned at
low momenta from becoming too strong at the cutoff momentum. \ Further details
will be discussed in a forthcoming publication. \ We show results at leading
order (LO), next-to-leading order (NLO), next-to-leading order with
isospin-breaking and electromagnetic corrections (IB + EM), and
next-to-next-to-leading order (NNLO). \ We follow the power counting scheme
used in Ref.~\cite{21}, and there is no additional isospin-breaking and
electromagnetic corrections at NNLO. \ All energies are in units of MeV. \ For
comparison we also give the experimentally observed energies.\begin{table}[tb]
\caption{Lattice results for the ground state energies for $^{4}$He, $^{8}$Be,
and $^{12}$C. \ For comparison we also exhibit the experimentally observed
energies. \ All energies are in units of MeV.}
\begin{tabular}
[c]{|l||c|c|c|}\hline
& $^{4}$He & $^{8}$Be & $^{12}$C\\\hline\hline
LO [$O(Q^{0})$] & $-24.8(2)$ & $-60.9(7)$ & $-110(2)$\\\hline
NLO [$O(Q^{2})$] & $-24.7(2)$ & $-60(2)$ & $-93(3)$\\\hline
IB + EM [$O(Q^{2})$] & $-23.8(2)$ & $-55(2)$ & $-85(3)$\\\hline
NNLO [$O(Q^{3})$] & $-28.4(3)$ & $-58(2)$ & $-91(3)$\\\hline
Experiment & $-28.30$ & $-56.50$ & $-92.16$\\\hline
\end{tabular}
\label{ground_states}%
\end{table}\ These calculations as well as all other results presented here
use lattice spacing $a=1.97~$fm and time step $a_{t}=1.32~$fm. \ To simplify
unit conversions we are using units where $\hbar$ and $c$, the speed of light,
are set equal to $1$. \ The error bars in Table~I are one standard deviation
estimates which include both Monte Carlo statistical errors and uncertainties
due to extrapolation at large Euclidean time. \ For each simulation we have
collected data from $2048$ processors each generating about $300$ independent
lattice configurations. \ In the case of $^{12}$C, these configurations are
stored on disk and used for the analysis of excited states described later.

For $^{4}$He the periodic cube length is $L=9.9~$fm, while the system size for
the $^{8}$Be and $^{12}$C calculations are each $11.8~$fm. \ By probing the
two-nucleon spatial correlations for each nucleus, we conclude that the finite
size corrections are smaller than the combined statistical and extrapolation
error bars. \ Since the lattice EFT calculations are based upon an expansion
in powers of momentum, the size of corrections from $O(Q^{0})$ to $O(Q^{2})$
and from $O(Q^{2})$ to $O(Q^{3})$ give an estimate of systematic errors due to
omitted terms at $O(Q^{4})$ and higher. \ We have used the experimentally
observed $^{4}$He energy to set one of the unknown three-nucleon interaction
coefficients at NNLO commonly known in the literature as $c_{D}$. \ However,
the results for $^{8}$Be and $^{12}$C are predictions without free parameters,
and the results at NNLO are in agreement with experimental values.

In order to compute the low-lying excited states of carbon-12, we generalize
the Euclidean time projection method to a multi-channel calculation. \ We
apply the exponential operator $e^{-Ht}$ to $24$ single-nucleon standing waves
in the periodic cube. From these standing waves we build initial states
consisting of $6$ protons and $6$ neutrons each and extract four orthogonal
energy levels with the desired quantum properties. \ All four have even parity
and total momentum equal to zero. \ Three states have $z$-axis component of
angular momentum, $J_{z}$, equal to $0$, and one has $J_{z}$ equal to $2$.
\ We note that the lattice discretization of space and periodic boundaries
reduce the full rotational group to a cubic subgroup. \ As a consequence only
$90$-degree rotations along axes are exact symmetries. \ This complicates the
identification of spin states. \ However the degeneracy or non-degeneracy of
energy levels for $J_{z}=0$ and $J_{z}=2$ allows one to distinguish between
spinless states and spin-2 states. \ We use the spectroscopic notation
$J_{n}^{\pi}$, where $J$ is the total spin, $\pi$ denotes parity, and $n$
labels the excitation starting from $1$ for the lowest level. \ In this
notation the ground state is $0_{1}^{+}$, the Hoyle state is $0_{2}^{+}$, and
the lowest spin-2 state is $2_{1}^{+}$.

\begin{figure}[ptb]%
\centering
\includegraphics[
height=2.0271in,
width=1.9873in
]%
{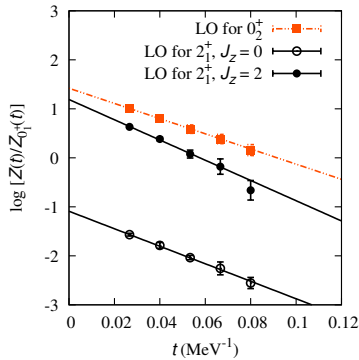}%
\caption{Extraction of the excited states of $^{12}$C from the time dependence
of the projection amplitude at LO. \ The slope of the logarithm of
$Z(t)/Z_{0_{1}^{+}}(t)$ at large $t$ determines the energy relative to the
ground state.}%
\label{carbon12_3jz0_1jz2_lo}%
\end{figure}

In Table~\ref{excited states} we show results for the low-lying excited states
of $^{12}$C at leading order (LO), next-to-leading order (NLO),
next-to-leading order with isospin-breaking and electromagnetic corrections
(IB + EM), and next-to-next-to-leading order (NNLO). \ All energies are in
units of MeV. For comparison we list the experimentally observed energies.
\ As before the error bars in Table~\ref{excited states} are one standard
deviation estimates which include both Monte Carlo statistical errors and
uncertainties due to extrapolation at large Euclidean time. \ Systematic
errors\ due to omitted higher-order interactions can be estimated from the
size of corrections from $O(Q^{0})$ to $O(Q^{2})$ and from $O(Q^{2})$ to
$O(Q^{3})$. \ In Fig.~\ref{carbon12_3jz0_1jz2_lo} we show lattice results used
to extract the excited state energies at leading order. \ For each excited
state we plot the logarithm of the ratio of the projection amplitudes,
$Z(t)/Z_{0_{1}^{+}}(t)$, at leading order. \ $Z_{0_{1}^{+}}(t)$ is the ground
state projection amplitude, and the slope of the logarithmic function at large
$t$ gives the energy difference between the ground state and the excited state.

\begin{table}[tb]
\caption{Lattice results for the low-lying excited states of $^{12}$C. \ For
comparison the experimentally observed energies are shown. \ All energies are
in units of MeV.}
\begin{tabular}
[c]{|l||c|c|c|}\hline
& $0_{2}^{+}$ & $2_{1}^{+}$, $J_{z}=0$ & $2_{1}^{+}$, $J_{z}=2$\\\hline\hline
LO [$O(Q^{0})$] & $-94(2)$ & $-92(2)$ & $-89(2)$\\\hline
NLO [$O(Q^{2})$] & $-82(3)$ & $-87(3)$ & $-85(3)$\\\hline
IB + EM [$O(Q^{2})$] & $-74(3)$ & $-80(3)$ & $-78(3)$\\\hline
NNLO [$O(Q^{3})$] & $-85(3)$ & $-88(3)$ & $-90(4)$\\\hline
Experiment & $-84.51$ & \multicolumn{2}{|c|}{$-87.72$}\\\hline
\end{tabular}
\label{excited states}%
\end{table}%

\begin{figure}[ptb]%
\centering
\includegraphics[
height=2.0254in,
width=2.5443in
]%
{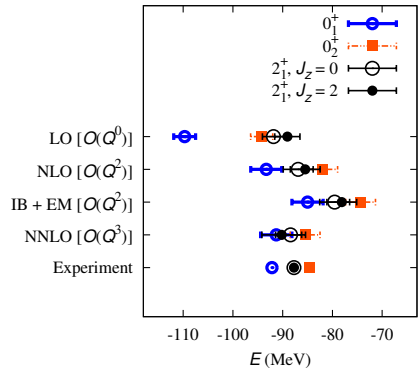}%
\caption{ Summary of lattice results for the carbon-12 spectrum and comparison
with the experimental values. For each order in chiral EFT labelled on the
left, results are shown for the ground state (blue circles), Hoyle state (red
squares), and the $J_{z}=0$ (open black circles) and $J_{z}=2$ (filled black
circles) projections of the spin-2 state.}%
\label{carbon12_summary_horizontal}%
\end{figure}

As seen in Table~\ref{excited states} and summarized in
Fig.~\ref{carbon12_summary_horizontal}, the NNLO results for the Hoyle state
and spin-2 state are in agreement with the experimental values. \ While the
ground state and spin-2 state have been calculated in other studies
\cite{10,11,13}, these results are the first \textit{ab initio} calculations
of the Hoyle state with an energy close to the phenomenologically important
$^{8}$Be-alpha threshold. \ Experimentally the $^{8}$Be-alpha threshold is at
$-84.80$~MeV, and the lattice determination at NNLO gives $-86(2)$~MeV. \ We
also note the energy level crossing involving the Hoyle state and the spin-2
state. \ The Hoyle state is lower in energy at LO but higher at NLO. \ One of
the main characteristics of the NLO interactions is to increase the repulsion
between nucleons at short distances. \ This has the effect of decreasing the
binding strength of the spinless states relative to higher-spin states. \ We
note the $17$~MeV reduction in the ground state binding energy and $12$~MeV
reduction for the Hoyle state while less than half as much binding
correction\ for the spin-2 state. \ This degree of freedom in the energy
spectrum suggests that at least some fine-tuning of parameters is needed to
set the Hoyle state energy near the $^{8}$Be-alpha threshold. \ It would be
very interesting to understand which fundamental parameters in nature control
this fine-tuning. \ At the most fundamental level there are only a few such
parameters, one of the most interesting being the masses of the up and down
quarks \cite{23,24}.%

\begin{figure}[ptb]%
\centering
\includegraphics[
height=2.2477in,
width=2.8046in
]%
{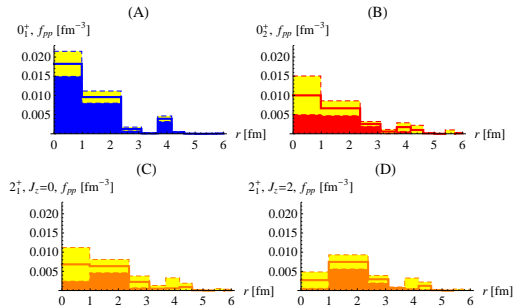}%
\caption{The radial distribution function $f_{pp}(r)$ for the ground state
(A), Hoyle state (B), and in the $J_{z}=0$ (C) and $J_{z}=2$ (D) projections
of the spin-2 state. \ The yellow bands denote error bars.}%
\label{radial_distribution}%
\end{figure}

Our comments on the binding energies at LO\ would also suggest that the
nuclear wavefunctions at LO are probably somewhat too compact for the spinless
states. \ We check for this explicitly by computing the proton-proton radial
distribution function $f_{pp}(r)$. \ Using any given proton as a reference
point, the function $f_{pp}(r)$ is proportional to the probability of finding
a second proton at a distance $r$. \ For macroscopic liquids the radial
distribution function is normalized to $1$ at asymptotically large distances.
\ In our finite system we instead normalize the integral of $f_{pp}(r)$ over
all space to equal $1-Z^{-1}$, where $Z$ is the total number of protons. \ In
Fig.~\ref{radial_distribution} we show the radial distribution function
$f_{pp}(r)$ at Euclidean time $t=0.08$~MeV$^{-1}$ for the ground state (A),
Hoyle state (B), and the $J_{z}=0$ (C) and $J_{z}=2$ (D) projections of the
spin-2 state. \ The yellow bands denote one standard deviation error bars.%

The ground state is very compact with a large central core. \ The Hoyle state
and spin-2 state look qualitatively similar, though the Hoyle state has a
slightly larger central core. \ A secondary maximum near $r\simeq4\,$fm is
visible in the ground state and each of the excited states. \ This secondary
maximum seems to arise from configurations where three alpha clusters are
arranged approximately linearly. \ More calculations are planned to confirm
whether this configuration is physically important or just a lattice artifact.

It is straightforward to compute the root-mean-square charge radius from the
second moment of $f_{pp}(r)$. \ We include the charge radius of the proton,
$0.84$ fm~\cite{25}, by adding it in quadrature. \ At LO\ we obtain a charge
radius of $2.04(2)~$fm for the ground state, $2.4(1)~$fm for the Hoyle state,
and $2.6(1)\,$fm and $2.4(2)\,$fm for $J_{z}=0$ and $J_{z}=2$ projections of
the spin-2 state. \ The experimentally observed charge radius for the ground
state is $2.47(2)\,$fm~\cite{26}. \ As expected the ground state wavefunction
at LO is too small by a proportion similar to the overbinding in energy. \ The
radius for the ground state and Hoyle state should increase significantly when
the NLO\ corrections are included. \ One expects some correction due to the
finite-volume periodic boundary. \ At LO the tail of the radial distribution
function suggests that this is a rather small effect for $L=11.8$ fm. \ Higher
order corrections to the radial distribution function, charge radii, as well
as electromagnetic transition strengths are currently under investigation and
will be discussed in a future publication. \ 

In summary we have presented \textit{ab initio} calculations of the low-lying
states of carbon-12 using lattice effective field theory. \ In addition to the
ground state and excited spin-2 state, we find a resonance with spin zero and
positive parity at $-85(3)$~MeV which appears to be the Hoyle state. \ Much
more work is needed and planned, including calculations at smaller lattice
spacings. \ But these lattice calculations provide a new opening towards
understanding the physics of this unique state and may also prove useful for
the study of other nuclear reactions relevant to the element synthesis in
stars.\medskip

{\small \noindent\textbf{Acknowledgements}~~ We thank Steven Weinberg for
useful communications. \ Financial support acknowledged from the DFG (SFB/TR
16), Helmholtz Association (VH-VI-231), BMBF\ (grant 06BN9006), U.S. DOE
(DE-FG02-03ER41260), EU HadronPhysics2 project \textquotedblleft Study of
strongly interacting matter\textquotedblright,\ and ERC project 259218
NUCLEAREFT. \ Computational resources provided by the J\"{u}lich
Supercomputing Centre at the Forschungszentrum J\"{u}lich. }

\end{document}